# Heavy Flavor Physics through e-Science

**Kihyeon Cho and Hyunwoo Kim**

e-Science Applications Research and Development Team,

Korea Institute of Science and Technology Information, Daejeon 305-806

Heavy flavor physics is an important element in understanding the nature of physics. The accurate knowledge of properties of heavy flavor physics plays an essential role for the determination of the Cabibbo-Kobayashi-Maskawa (CKM) matrix. Asymmetric-energy $e^+e^-$ B factories (BaBar and Belle) run their operation and will upgrade B factories to become super Belle. The size of available B meson samples will be dramatically increased. Also the data size of Tevatron experiments (CDF, D0) are on the order of PetaByte. Therefore we use new concept of e-Science for heavy flavor physics. This concept is about studying heavy flavor physics anytime and anywhere even if we are not on-site of accelerator laboratories and data size is immense. The component of this concept is data production, data processing and data analysis anytime and anywhere. We apply this concept to current CDF experiment at Tevatron. We will expand this concept to Super Belle and LHC (Large Hadron Collider) experiments which will achieve an accuracy of measurements in the next decades.

Fax: +82-42-869-0789

# I. INTRODUCTION

Heavy flavor physics is an important element in understanding the nature of particle physics. There are three known generation of quark doublets. (u,d), (c,s), (t,b). However, the origins of families are unknown in the standard model. Only the charged-current electroweak interaction can change flavors in the standard model. Electroweak eigenstates are not mass eigenstates, which introduced CKM (Cabibbo-Kobayashi-Maskawa) matrix [1]. There are only the standard model connections between generations. CKM matrix elements are fundamental parameters of the standard model. They cannot be predicted theoretically but can be measured by experiments. CKM matrix has three rotation parameters and one phase parameter. Even if the non-zero phase parameter implies CPV (Charge Parity Violation) in flavor transitions, its origin is still one of the mysteries of elementary particle physics. The CPV can occur when multiple B decay amplitude interfere each other. B system is uniquely situated for CPV studies due to mixing, long lifetime, large prediction cross section, rich decay set and heavy quarks, which is theoretically accessible.

Therefore, the precision measurement of heavy flavor physics provides confirmation of CKM theory. However, the standard model leaves many questions about the flavor sector unanswered such as the origin of generation and masses, a mixing and disappearance of antimatter. Core properties of weak interactions provide the parameters not predicted within the standard model. The standard model measurement provides the foundation of theory while a deviation may give a hint on new physics beyond the standard model. The unitary triangle represents a graphical expression of unitary conditions. CKM unitary violation would imply new physics. We may test the standard model and CKM by over-constraining angles and sides. Quantum effect beyond the standard model has very small probability such as b→s transition. Only huge B data would provide the information on new physics.

Therefore, in order to handle more data and more collaboration efficiently, it is time to study heavy flavor physics through e-Science. We apply this concept to CDF (Collider Detector at Fermilab) experiment and show the possible applications for future LHC and Super Belle experiments.

## II. HEAVY FLAVOR PHYSICS EXPERIMENTS

*1. Overview*

Current heavy flavor physics experiments in lepton colliders are BESII, CLEOc, Belle and BaBar. These experiments can achieve excellent photon resolutions due to good calorimeters [2] and can provide exact information on missing energy by energy and momentum conservations. Recently, Belle reports a difference between direct CP (Charge Parity) violation in charged and neutral B meson decays [3].

In the meantime, heavy flavor physics experiments in hadron colliders provide large data set and heavier b hadrons. The cross section for proton and anti-proton is $\sigma(p\bar{p} \to b\bar{b}) \sim 150 \mu b$ at 2 TeV while $\sigma(e^+ e^- \to b\bar{b}) = 1 nb$ at Y(4S). Tevatron also produces heavy b such as $\Lambda_b$ and $B_s$. Therefore, we call the Tevatron as the full service of B factory. The results from lepton collider and hadron collider are complementary. At the Tevatron, B hadrons are mostly produced in pairs. The main $b\bar{b}$ production mechanism is flavor creation through gluon fusion. Due to the excellent performance of the Tevatron the CDF experiment has recorded 4 fb$^{-1}$ data by September 2008. However, the Tevatron $b\bar{b}$ cross section is orders of magnitude smaller than the total inelastic cross section of around 50 mb. For this reason, CDF employ triggers that select events with signatures specific to various B decays. We use the dimuon trigger, geared towards $B \to J/\psi X$ decays and the displaced track trigger used for measurement of $B_s$ lifetime [4]. The heavy flavor physics results show the precision measurement of mass and lifetime of $B_c$ mesons, $B_s$ lifetime and width difference and CPV (CP violation) in $B_s$ system.

The highlighted heavy flavor physics results from CDF experiment are 1) observation of $B_s$ with $\Delta m_s$ = 17.77 ± 0.10 (stat.) ± 0.07(sys.) [5], 2) observation of new baryon states $\Sigma_b$ and $\Xi_b$, 3) single top quark observation (4.4 sigma) using 2.7 fb$^{-1}$ data that the cross section of single top quark is 2.2 pb [6], 4) measurement of $\sin(2\beta_s)$, 5) precision measurement of top mass, $M_{top\_CDF}$ = 172.4 ± 1.0 ± 1.3GeV/$c^2$ [7], 6) observation of new charmless $b \rightarrow hh$ states, and 7) observation of $D^0 - \overline{D^0}$ mixing. CDF has also presented recent results on $B_c$ and $B_s$ meson properties. The best $B_c$ mass measurement was performed by CDF in fully reconstructed $B_c \rightarrow J/\psi\pi$ decays. The measurements of the $B_c$ lifetime in semileptonic decays agree with the measurements from the D0 experiment with similar precision. These measurements provide useful information for the unitarity angles that can be used to improve theoretical models used to study heavy mesons.

Figure 1 shows the current measurement of unitarity triangle by both lepton collider experiments (BESII, CLEOc, BaBar and Belle) and hadron collider experiments (CDF and D0) [8].

## 2. Heavy flavor physics experiments in the next decades

High-energy physics in the next decades will be focused on three topics. The first topic is energy frontier experiment by LHC and ILC (International Linear Collider) colliders to find and probe Higgs, supersymmetry, dark matter and new understanding of space-time. The second topic is lepton physics experiment by neutrino experiment to study neutrino lepton flavor violation, tau lepton flavor violation to find neutrino mixing and masses, and lepton number non-conservation. The third topic is heavy flavor physics by super B factory to find CP asymmetry, baryogenesis, left-right symmetry and new sources.

Heavy flavor physics by LHC (Large Hardon Collider) experiments - CMS (Compact Muon Solenoid), ATLAS (A Toroid LHC ApparatuS) and LHCb - are energy frontier experiments. For heavy flavor physics at CMS and ATLAS, some benchmark analyses are 1) cross section for bottom, charm, and quarkonia, 2) correlation studies, 3) quarkonia analysis: polarization, production mechanism, 4) lifetime and properties b hadrons: $B^+$, $B_d$, $B_s$, $B_c$, and $\Lambda_b$, 5) $B_s$ oscillations and CP violation, 6) FCNC (Flavor Changing Neutral Current) rare decays: $B_d \to K^{*0}\mu^+\mu^-$, $B_s \to \phi\mu^+\mu^-$ and $B^+ \to K^+\mu^+\mu^-$, 7) FCNC very rare decay $B_s \to \mu^+\mu^-$ or $B_d \to \mu^+\mu^-$ and 8) lepton flavor violation such as $\tau \to 3\mu$ [9]. For these analyses, triggers in CMS and ATLAS are used from excellent muon system, trackers and large acceptance. In CMS and ATLAS experiments, most analyses are focused on muons [10].

For heavy flavor physics at the LHCb experiment [11, 12], the luminosity is $2 \sim 5 \times 10^{32} cm^{-2} s^{-1}$ and the cross section of $b\bar{b}$ is $500 \mu b$. This produces lots of events up to 2 fb$^{-1}$ for the first year. The LCHb experiment is a heavy flavor precision experiment searching for new physics in CP violation and rare decays. The LHCb detector is a unique forward detector as opposed to other experiments using barrel detector. The LHCb experiment is designed to study CP violation in the b-quark sector at the LHC and expand the current studies underway at the B-factories (Babar, Belle) and at the Tevatron (CDF, D0). The LHCb opens the opportunity to study B-hadrons that cannot be produced at current B-factories, and the energy of 14 TeV, much higher than that of the Tevatron, allows an abundant production of B particles ($10^5$ particles/s at the nominal luminosity) [13]. The $b\bar{b}$ production cross section is 2 orders of magnitude smaller than the total cross section visible in the detector, and the decay modes of the b hadrons that are of interest for CP violation studies have very low visible branching fractions, typically smaller than 10$^{-4}$. Hence a very selective and sophisticated trigger is needed. LHCb is planning to operate a 3-level trigger system to select the events of interest [13].

Heavy flavor physics at super Belle plans to start in 2012. The integrated luminosity will be 70 ab$^{-1}$ between 2012 and 2020. For the Super Belle experiment two key measurements are non-SM (Standard Model) CP violation from B meson decay and lepton flavor violation in tau decay. In detail, the experiment will provide 1) non-SM (Standard Model) CP phase from high precision b→s penguin studies, 2) charged Higgs from searches in $B^+ \to \tau^+\nu$ and $B \to D^{(*)}\tau^+\nu$, 3) non-SM right-handed current from $B \to K^*\gamma$ CPV, 4) Inclusive measurements from $b \to s\gamma$, $b \to d\gamma$, $b \to sl^+l^-$ and $V_{ub}$, 5) understanding of loop vs. tree diagram from high precision unitarity triangle measurement, 6) lepton flavor violation from searches in high statistics tau decays and 7) new physics searches in up-quark sector from CPV in $D^0 - \overline{D^0}$ mixing [14]. From Super Belle experiment we may find new particles such as fourth quark states, endless list for rare B decays, $B_S$ at $Y(5S)$ and more D decays. Now it is time to look more closely at the physics states. We need more focus on early stage of upgraded B factory at KEK to maximize physics output with 3 or 5 $ab^{-1}$ data. The 10 $ab^{-1}$ data would tell us the direction of flavor physics which is goal of current roadmap. The 50 $ab^{-1}$ data would allow us to study flavor physics beyond standard model towards the systematics and theory limits [14].

## 3. Heavy flavor physics experiments in the LHC era

Now that LHC turns on, LHC experiment by high transverse momentum ($p_T$) will provide a unique effort towards the high-energy frontier to determine energy scale of new physics. Meanwhile, collective efforts toward the high-intensity frontier will provide flavor physics by rare K decays, universality tests in B and K, CPV in the $B_S$ system, improved CKM fits, rare B decays, LFV (Lepton Flavor Violation) in muon and tau decays and g-2 to determine the flavor structure of new physics [15].

## 4. The trend of heavy flavor physics experiments

Since the invention of the cyclotron by Ernest Orlando Lawrence at University of California, Berkeley in 1930 [16], accelerators and physics experiments have changed a lot. We explain the recent trend of heavy flavor physics experiments from data production, data processing and data analysis point of view.

From data production point of view, now the center of mass energy is higher than the past ranging from 10.56 GeV of $Y(4S)$ of B factory experiments to 14 TeV of LHC experiments.

From data processing point of view the cross section is also higher than the past from lepton collider ($e^+e^-$) to hadron collider (p-p), which produces more data. Table 1 shows the size of production data for heavy flavor physics experiments. The cross section of LHC experiment is $\sigma(pp \to b\bar{b}) \sim 500 \mu b$ at $\sqrt{s}=14$ TeV while that of B factories is $\sigma(e^+ e^- \to b\bar{b}) = 1nb$ at Y(4S). Therefore, LHC produces 500,000 times more data than that of B factory experiments. The size of the data is on the order of PetaByte per year. Meantime, the Belle experiment has measured fundamental parameters by using a large sample that corresponds to an integrated luminosity of 0.9 ab$^{-1}$, or 900 million B anti-B meson pairs, and has consequently verified the framework of the standard model since its operation in 1999. Now the Belle collaboration is proposing the Super Belle experiment from 2012, a project that can detect new phenomena that arise from physics beyond the standard model by using a much larger amount of data as 70 ab$^{-1}$ or 70 billion B anti-B meson pairs.

From data analysis collaboration point of view, we have more collaboration. CLEO experiment needed around 200 physicists, whereas now 2,000 physicists work for CMS or ALICE experiments. The number of collaboration is dramatically increased. Therefore, in order to cope with more data and more collaboration as well as higher energy and higher cross section, it is time to use e-Science paradigm for heavy flavor physics.

### III. E-SCIENCE

*1. The definition of e-Science*

Thousand years ago, science was an experimental science to describe natural phenomena. For the last few hundred years, science has been a theoretical science such as Newton's laws and Maxwell's equations. For the last few decades, science has been a computational science for simulation of complex phenomena. Today, science becomes e-Science which is data-centric science to unify theory, experiment and simulation. Now science includes many challenging problems that require large resources, particularly knowledge from many disciplines [17]. e-Science is a new research paradigm for science, which is computationally intensive and carried out in highly distributed network environments [18]. The e-Science uses immense data sets that require grid technology [18].

*2. The goal of e-Science*

The goal of e-Science is to do any research "anytime and anywhere." High-energy physics experiments are usually conducted at major accelerator sites (on-site) where experimentalists perform detector design and construction, signal processing, data acquisition, and data analysis on a large scale [19]. Therefore, the goal of e-Science for high-energy physics is to study high energy physics "anytime and anywhere" even if we are not at accelerator laboratories (on-site). High-energy physics requires a particularly well-developed e-Science infrastructure due to its need for adequate computing facilities for the analysis of results and the storage of data [18]. To perform computing at the required high-energy physics scale, we need data grid technology.

*3. The components of e-Science*

As shown in Figure 2, the components of e-Science for high energy physics include 1) data production, 2) data processing, and 3) data analysis collaboration that can be accessed any time and anywhere even if we are not on-site.

First, data production is to take both on-line shift and off-line shift anywhere. On-line shift is taken through remote control room and off-line shift is taken through data handling system. Second, data processing is to process data by using a high-energy physics data grid. The objective of a high-energy physics data grid is to construct a system to manage and process high-energy physics data and to support the high-energy physics community [20]. Third, data analysis is for collaborations around world to analyze and publish the results by using collaborative environments. We apply this concept to CDF experiment at Fermilab.

# IV. RESULTS

We show an example of e-Science by the CDF experiment. Table 2 shows the components of e-Science compared to Fermilab (on-site). In Fermilab (on-site), we have B0 trailer in which we take shifts for DAQ (Data Acquisition). The first component is data production. In Fermilab, there is Feynman computing center to process data. The second component is data processing. There is a conference room to discuss analysis. The third component is data analysis collaboration. We can do these three components remotely (off-site).

For data production, we have taken remote shift both on-line and off-line. For the on-line shift, we have constructed a remote control room for a CO (Consumer Operation) operation room at the KISTI. The CDF control room at Fermilab consists of many sections. One of them is monitoring sections to check the quality of data. We call it CO, which does not affect the control and the data acquisition directly. Everything on the CDF CO desktop is available at remote sites. The status and logs of 'Event Display', 'Consumer Slide', 'Consumer Monitors' and 'Detector Check List' are done through a web site. A 'Consumer Controller' can be sent to the remote site from CDF control room at Fermilab [19]. Communication with shift crews at the CDF control room is done by using Polycom system. For off-line shift we have taken SAM (Sequential Access through Meta-data) data handling shifts [21].

For data processing, we use the grid technology. From the technical point of view, it should have scalability [22]. It consists of O(100) sites, O(100k) CPU, O(10PByte) disk, O(100k) jobs/data, O(10M) files, O(10Gb/s) transfer, O(10k) users and O(100) VO (Virtual Organization). We still need high operation cost for man power. From the collaborative point of view, it should have accessibility. In other words, it should be equivalent to all collaborations and it must have global infrastructure for communication. From political and financial point of view, it should have visibility. It should be recognized in institute and country. Even if we must provide our own resources, the grid technology can now be a common solution for future high energy physics programs. The CDF experiment shows one of good examples. CDF is using several computing processing systems, such as CAF (Central Analysis Farm) [23], DCAF (Decentralized CDF Analysis Farm) [20] and grid systems. Table 3 shows the progress of the CDF grid [19]. Based on the grid concept [24], CDF has constructed a CDF VO. Moreover, a significant fraction of these resources is expected to be available to CDF during the LHC era. The movement to a grid at the CDF experiment is in synchronization with the worldwide trend for high-energy physics experiments. We have made a federation of the LCG and OSG (Open Science Grid) farms at the AS (Academia Sinica) in Taiwan, the LCG farm at the KISTI in Korea and the LCG farm at the University of Tsukuba in Japan. We call this federation of grid farms as 'Pacific CAF'.

For data analysis, we have constructed the EVO (Enabling Virtual Organization) servers at the KISTI. When users in Korea use the EVO servers at the KISTI, the routing time is reduced by 60 msec without the congestion of the network inside of USA, which gives very stable research environment [19].

For the outlook of e-Science, we have final dataset of around 500 million B decays from BaBar experiment. Belle experiment continues operations to order of 1,000 million decays. Therefore, we will have final results from combined B factories. We expect to double the data set at the Tevatron. Therefore, high-precision measurement is around the corner. We will also have data set of LHC experiments and Super B factories in Japan near future.

## V. CONCLUSION

Heavy flavor physics has an important role for CP violation and decay mechanisms. Due to more data and more collaboration as well as higher energy and higher cross section, it is time to study heavy flavor physics through e-Science. We apply this concept to CDF experiment and show the possible applications for future experiment of LHC and Super Belle experiments.

## ACKNOWLEDGMENTS

We would like to thank Minho Jeung (Korea Institute of Science and Technology Information) for data processing at LCG (LHC Computing Grid) farm.

Table 1. The size of production data for heavy flavor physics experiments.

| Collider | Current experiment | | Next experiment | |
|---|---|---|---|---|
| | Experiment | Raw data size | Experiment | Raw data size |
| Lepton Collider ($e^+e^-$) | Belle (1999-present) BaBar (1999-2008) | ~1 PByte (0.9 ab$^{-1}$) ~0.5PByte(0.5 ab$^{-1}$) | Super Belle (2012- ) | 5~10 PByte/year |
| Hadron Collider ($p-\bar{p}/p-p$) | CDF (2001-present) D0 (2001-present) | ~2 PByte (4.0 fb$^{-1}$) ~2 PByte (4.0 fb$^{-1}$) | CMS (2008- ) ATLAS (2008-) LHCb (2008- ) | 5~10 PByte/year 5~10PByte/year 0.2~1 PByte/year |

Table 2. The components of e-Science compared to typical science at Fermilab.

| Component | Typical Science | e-Science |
|---|---|---|
| Site (Location) | On-Site (Fermilab, USA) | Off-Site (KISTI, Korea) |
| Data Analysis | Conference Room | EVO (Enabling Virtual Organization) |
| Data Processing | CAF (Central Analysis Farm) | Pacific CAF (CDF Analysis Farm) |
| Data Production - On-line shift - Off-line shift | CDF Control Room at Fermilab Off-line shift at Fermilab | Remote Control Room at KISTI SAM Data Handling shift at KISTI |

Table 3. The progress of CDF Grid [19].

| Name | Starting date | Grid middleware | Job scheduling | Content | Site |
|---|---|---|---|---|---|
| CAF (Central Analysis Farm) | 2001 | - | Condor | Cluster farm inside Fermilab | USA (Fermilab) |
| DCAF (Decentralized CDF Analysis Farm) | 2003 | - | Condor | Cluster farm outside Fermilab | Korea, Japan, Taiwan, Spain (Barcelona, Cantabria), USA (Rutgers, San Diego), Canada, France, |
| Grid CAF (CDF Analysis Farm) | 2006 | LCG / OSG | Resource Broker + Condor | Grid farm | North America CAF European CAF Pacific CAF |
| CGCC (CDF Grid Computing Center) | 2008 | LCG | Resource Broker + Condor | Big Grid farm + Storage | Korea (KISTI) France (IN2P3) Italy (CNAF) |

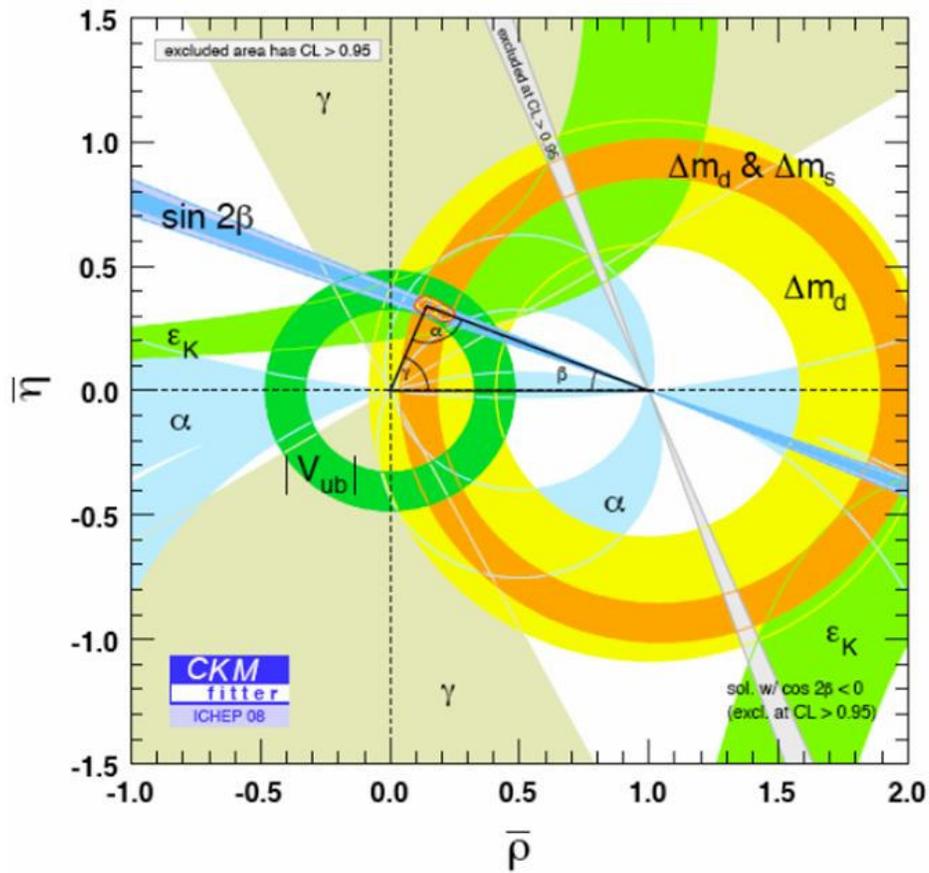

Fig. 1 The current measurement of the unitarity triangle by both lepton collider experiments (BESII, CLEOc, BaBar and Belle) and hadron collider experiments (CDF and D0) [8].

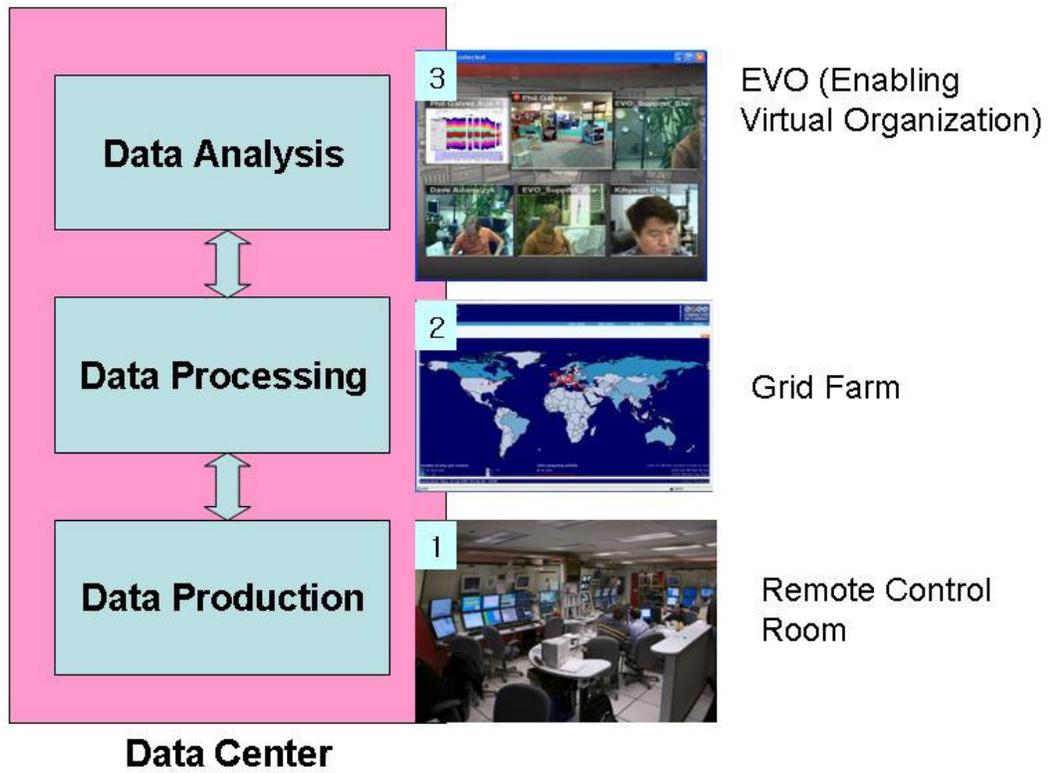

Fig. 2. The typical components of e-Science for high-energy physics.